One-sign order parameter in iron based superconductor

S.V. Borisenko[1], V. B. Zabolotnyy[1], A. A. Kordyuk[1,2], D. V. Evtushinsky[1], T. K. Kim[1,3], I. V. Morozov[4], R. Follath[5], B. Büchner[1]

[1]Leibniz-Institute for Solid State Research, IFW-Dresden, D-01171 Dresden, Germany

[2]Institute for Metal Physics, 03142 Kyiv, Ukraine

[3]Diamond Light Source Ltd., Didcot, Oxfordshire, OX11 0DE, United Kingdom

[4]Moscow State University, 119991 Moscow, Russia

[5]Helmholtz-Zentrum Berlin, BESSY, D-12489 Berlin, Germany

**The onset of superconductivity at the transition temperature is marked by the onset of order, which is characterized by an energy gap. Most models of the iron-based superconductors find a sign-changing (s±) order parameter [1-6], with the physical implication that pairing is driven by spin fluctuations. Recent work, however, has indicated that LiFeAs has a simple isotropic order parameter [7-9] and spin fluctuations are not necessary [7,10], contrary to the models [1-6]. The strength of the spin fluctuations has been controversial [11,12], meaning that the mechanism of superconductivity cannot as yet be determined. Here we report the momentum dependence of the superconducting energy gap, where we find an anisotropy that rules out coupling through spin fluctuations and the sign change. The results instead suggest that orbital fluctuations assisted by phonons [13,14] are the best explanation for superconductivity.**

LiFeAs is a very remarkable member in the family of iron based superconductors (IBSC). While consisting of key FeAs structural blocks and loosing resistance already at 18K thus bearing all necessary and representative features of these novel materials [15, 16, 17], LiFeAs superconducts in its stoichiometric composition, is not magnetic [11, 12] and its surface electronic structure is the same as in the bulk [18]. Angle-resolved photoemission (ARPES) at ultra-low temperatures (~1K) (Ref. 14) offers a unique opportunity to directly determine the full size of the energy gap ($\Delta$) as a function of momentum, virtually in the ground state of a superconductor. Precise and direct determination of the energy gap becomes possible if the maximum energy position of the quasiparticle peak at Fermi momentum ($\mathbf{k}_F$) with respect to Fermi level ($E_F$) is determined (inset to Fig.1). LiFeAs is an ideal compound for such ARPES measurements --- all Fermi surfaces (locus of $\mathbf{k}_F$ at $E_F$) are well defined and separated in **k**-space (Fig.1a), no other gap or depletion of the spectral weight at $E_F$ (pseudogap) is present and $\mathbf{k}_F$ energy distribution curves (EDC) are remarkably sharp.

As has been demonstrated previously [7], the Fermi surface of LiFeAs consists of two hole-like and two electron-like sheets around the center and corners of the Brillouin zone (BZ) respectively (Fig.1a). We start with the determination of the energy gap corresponding to the large hole-like Fermi surface around Γ-point. This Fermi surface originates purely from in-plane $d_{xy}$ orbitals and therefore is 2D without any noticeable $k_z$ dispersion. Single dispersing features shown in Fig. 1 b,c represent the spectral function in the vicinity of the Fermi level at different momenta (A and B). Already visually, the distributions of intensity are not equivalent. Tracking the EDC's maxima close to

$E_F$ (Fig. 1 d) one clearly notices the difference: the energy gap is not the same in points A and B since the dispersion bends back at different distances from $E_F$ and this distance is equal to $\Delta$. This is a direct evidence for the gap anisotropy. We have determined the values of $\Delta$ for many $\mathbf{k}_F$ along the large Fermi surface and the result is shown in Fig.1 e,f (see Supplementary Information). An oscillating behavior, with the functional form $\Delta \sim \Delta_0 + \Delta_1 \cos 4\phi + ...$ is clearly seen. Even the presence of higher harmonics, $\cos 8\phi$ (~ 12%) and $\cos 12\phi$ (~ 20%) can be noticed as local extrema at $\phi=\pi n/4$ and asymmetric shape of global maxima and minima (see SI). The latter are oriented towards the sides and the corners of the BZ respectively.

In Fig. 2a we show a typical intensity distribution along one of the radial cuts from the corner of the BZ. This time two dispersing features, which supported electron-like FSs above $T_c$, are seen in spite of a finite three-dimensionality due to peculiar orbital composition (mostly $d_{xz,yz}$ but with finite admixture of $d_{xy}$) of these FS sheets (see SI). Panels b) and c) of Fig.2 show momentum dependences of both, the peak and leading edge positions for selected angles which define the direction of the radial cut (see inset to Fig. 2). From presented results one can see that the gap on the outer FS contour is smaller than on the inner one and that the gaps change in-phase, i.e. both increase when going from $\phi=0$ to the direction towards the $\Gamma$-point. This behavior is confirmed by the more detailed scanning of the angle and plotting the angular dependence for the two gaps (Fig. 2d). To investigate the gap function on a larger angular scale, we take advantage of the gap in-phase variations and consider in the following the averaged for both electron-like FSs gap. In such a way we increase the signal-to-noise ratio, otherwise significantly reduced by nearly degeneracy of FSs close to the crossings of two ellipses and by matrix-element effects. Large scans (Fig. 2e,f) recorded at different excitation energies (effectively, different $k_z$'s) again demonstrate approximately $\cos 4\phi$ behavior with global maxima of the gaps oriented towards the center of the BZ.

Finally, we determine the superconducting energy gap on the small FS at $\Gamma$-point. First of all, we note that the third, closest to $\Gamma$, hole-like $d_{xz,yz}$ feature strongly disperses towards the Fermi level but never crosses it, approaching as close as 10 meV at particular $k_z$. A detailed photon energy dependent study (not shown) has confirmed this and showed that the middle hole-like feature supports even smaller FS than it was initially thought [7]. The sizeable, BZ dependent intensity distribution seen in the vicinity of $\Gamma$-point is caused by the Van Hove singularity, but the actual FS crossing has been found only for certain $k_z$ intervals as is demonstrated by Fig. 3. Because of finite $k_z$-dispersion (of the order of 10 meV), the spectra will hardly be sensitive to the onset of superconductivity if the top of this band is situated more than 6 meV away from the Fermi level. It is the case, for instance, when using 20 eV and 50 eV photons (Fig. 3a). Nevertheless, higher temperature measurements (Fig. 3b) do indicate that the band reaches $E_F$ resulting in the so-called "shape resonance" [19]. For those $k_z$ the feature undergoes a noticeable transformation when crossing $T_c$ (Fig. 3b,c). The superconducting energy gap of the size of 6 meV opens and is in agreement with the tunneling spectroscopy and ARPES on Co-doped NaFeAs [20, 21, 22].

In Fig.4a we summarize our study by presenting the gap function schematically for the whole BZ. The largest gap (~6 meV) corresponds to the small hole-like FS at $\Gamma$-point. Along the large 2D hole-like FS the gap varies around ~3.4 meV roughly as 0.5eV*$\cos 4\phi$ meV being minimal at the direction towards the electron-like FS. The gap on the outer electron pocket is smaller than on the inner one and both vary around ~3.6 meV as 0.5eV*$\cos 4\phi$ having maximal values at the direction towards $\Gamma$-point. Since ARPES is not sensitive to the sign of the gap, we also sketch (Fig. 4b) the other possible gap function which would agree with our observations. We now examine to which extent the obtained gap structure corresponds to the popular s± spin-fluctuation scenario for superconductivity in iron pnictides [1-6]. There are two main possibilities for the realization of the s± scenario in terms of gap functions. It can be either nodal and have a functional form with dominant $\cos k_x + \cos k_y$ term or nodeless and have a functional form with dominant $\cos k_x * \cos k_y$ term, depending on the details of the pair interaction. In real space the former corresponds to the direct Fe-Fe exchange interaction while the latter to the next-nearest-neighbor exchange interaction via As atoms. The Fe-Fe distance in LiFeAs is one of the smallest among all families of pnictides and thus the first possibility is more likely. Moreover, many model calculations of the gap function unanimously predict nodal s± behavior for strongly electron doped materials [1-6]. This is natural since upon electron doping the size of hole-like FSs becomes negligible in comparison with the size of electron-like FSs, nesting is destroyed and spin-fluctuation mechanism can only be established when dominant interaction between electron-like pockets would change sign of the order parameter between them (in unfolded BZ). The system should adjust the magnitude of the angle-dependent, $\pm \cos 2\phi$ gap component along the two electron FSs to minimize the effect of the inter-electron-pocket repulsion [4]. LiFeAs, though being stoichiometric because of large $d_{xy}$ FS, should adopt this scenario for strongly electron-doped systems because the proportion between the $d_{xz,yz}$ originated FSs is exactly like this. Only one of the hole-like FSs with this orbital composition is present (the other one being completely below the $E_F$) whereas the electron-like pockets are very large with absolutely no sign of $(\pi, \pi)$-nesting [7]. These considerations inevitably imply the presence of nodes and anti-phase behavior of the gaps on electron-like FSs of LiFeAs (Fig. 4d), which is clearly not the case according to our experiment. Also, in the simplest case of $\cos k_x + \cos k_y$ order parameter one would expect the maximal gaps on the large hole-like FS to be oriented in the direction of electron pockets (Fig. 4d). Our results demonstrate just the opposite (Fig. 4a and Fig. 1e,f).

Let us assume that the unlikely for effectively electron doped IBSC scenario of nodeless s± order parameter takes place. Indeed, this would fix the in-phase/anti-phase discrepancy and the orientation of the maxima of the gap on the large hole pocket (Fig. 4c). However, the orientation of the extrema of the in-phase oscillating gaps on electron FSs would then be in conflict with the experimentally observed one (Fig. 4b). It is known that the orbital character of small portions of electron FSs oriented towards Γ-points is $d_{xy}$ and the interorbital pairing with $d_{xz,yz}$ states on hole FSs is weaker than the intraorbital pairing for the major parts of electron pockets. On the contrary, our data show that the gaps on these portions are maximal (Fig. 4a,b and Fig. 3e,f,g).

Another ingredient necessary for realization of spin-fluctuation scenario is apparently missing in LiFeAs. Recent neutron-scattering experiments on similarly prepared single crystals demonstrate that the strength of spin fluctuations in LiFeAs is an order of magnitude weaker than e.g. in Co-doped 211 system with hardly any evidence for the neutron resonance [11]. As a consequence, we were not able to find any evidence of electron-magnon coupling in LiFeAs contrary to the case of optimally doped 122 or exemplary case of cuprates [23]. Moreover, the detected energy spectrum of magnetic fluctuations does not contain any features that can explain the dispersion kinks in electronic spectrum observed earlier [10, 11, 12]. In Fig. 3d we show more evidence for strong electron-boson coupling. All typical energy scales derived in the present and previous [10] studies correspond to phonon modes [24], recently detected experimentally for the center of the BZ by Raman scattering [25] (see also SI). In spite of the noticeable renormalization (kinks) implying considerable electron-phonon coupling, the largest gap in LiFeAs is twice larger than in a weak coupling BCS scheme and its momentum dependence is very peculiar implying that conventional electron-phonon coupling mechanism is not operational in this material.

This brings us to an alternative approach to superconductivity in pnictides based on the orbital fluctuations model [13, 14]. A general consensus exists regarding the important role of orbital degrees of freedom played in physics of IBSC. Together with the possibility of Fe3d orbital ordering at structural transition [26], one of the most remarkable and robust experimental evidences for this importance is the universal bandwidth renormalization of the factor of 2-3 found in all families of pnictides and chalcogenides [7, 27, 28]. This renormalization is captured by e.g. DMFT calculations which take into account the Hund's rule coupling [29, 30, 31]. It is therefore not surprising that fluctuations of orbital order can drive the pairing. It was shown that for this interaction to be attractive, a moderate electron-phonon coupling should be present [13, 14].

The hallmark of the orbital-fluctuations-mediated-by-phonons scenario is the s++ order parameter, i.e. the superconducting gap of one sign for the whole BZ. According to the authors of Ref. 13, this superconducting state is likely to be realized exactly in electron-doped IBSC even without impurities (see Fig. 4c in Ref. 13). Taking into account the above remark about the classification of LiFeAs in terms of doping and directly determined gap function for all FSs, this is in perfect agreement with our data. The gap function in band representation calculated in Ref. 14 (Fig. 7) captures all the peculiarities of experimentally observed one: gaps on electron pockets oscillate in phase, their maxima are oriented towards Γ-point and even the orientation of the extrema on large hole-like $d_{xy}$ FS match the experimental observation.

While the details of the orbital fluctuations mechanism in LiFeAs are still to be understood, for instance, which exactly phonons are crucial for realization of this scenario [13, 14] or which role is played by the van Hove singularity, intraband nesting of $d_{xy}$ sheet or higher harmonics, implied by the shape of the gap function, our results clearly suggest that this interaction is a most promising alternative to conventional electron-phonon coupling and to conventional spin fluctuations.

**Acknowledgements** We acknowledge useful discussions with Andrey Chubukov, Ilya Eremin, Igor Mazin, Antonio Bianconi, Carsten Honerkamp, Christian Hess, Hidenori Takagi, Dima Efremov and Sergey Skornyakov . This work was supported by the DFG priority program SPP1458, Grants No. KN393/4, BO1912/2-1. I.M. acknowledges support from the RFBR-DFG (Project No. 10-03-91334).

**Competing Interests** The authors declare that they have no competing financial interests.

**Correspondence** Correspondence and requests for materials should be addressed to S. V. Borisenko

(S.Borisenko@ifw-dresden.de).

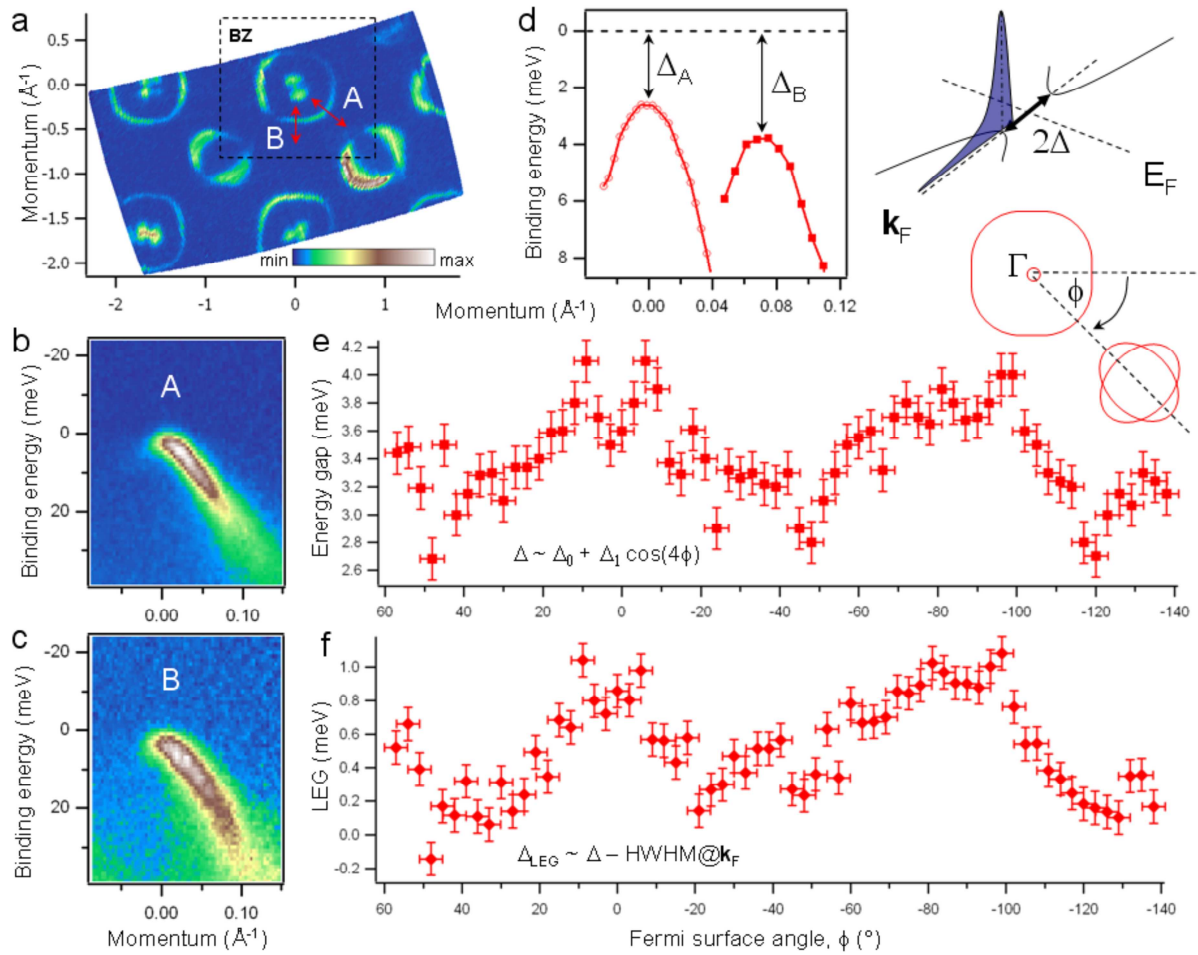

Figure 1. **Strong anisotropy of the superconducting gap. a)** Fermi surface map of LiFeAs. Large Fermi surface around center as well as crossed elliptical FSs around the corners are seen as closed features with intensity suppression along the high-symmetry directions because of matrix-element effects. Intensity distribution in the center of the BZ strongly depends on the BZ number and is due to the van Hove singularity below the Fermi level. The corresponding dispersing feature crosses the Fermi level only in very limited $k_z$ interval (see Fig. 3) thus resulting in a very small FS. Red arrows show the momentum location of the cuts A and B. **b,c)** Energy-momentum intensity distributions showing the crossings of the Fermi surface in points A and B. The typical bending back of the dispersion because of energy gap is seen. **d)** EDC dispersions from panels b and c. The gaps in points A and B are clearly different. **e)** Gap function from the peak positions of $k_F$-EDCs (see inset). **e)** Gap function from the positions of the EDC's leading edge midpoint, known as leading edge gap (LEG) and determined from the $k_F$-EDCs as energy position of a derivative's maximum.

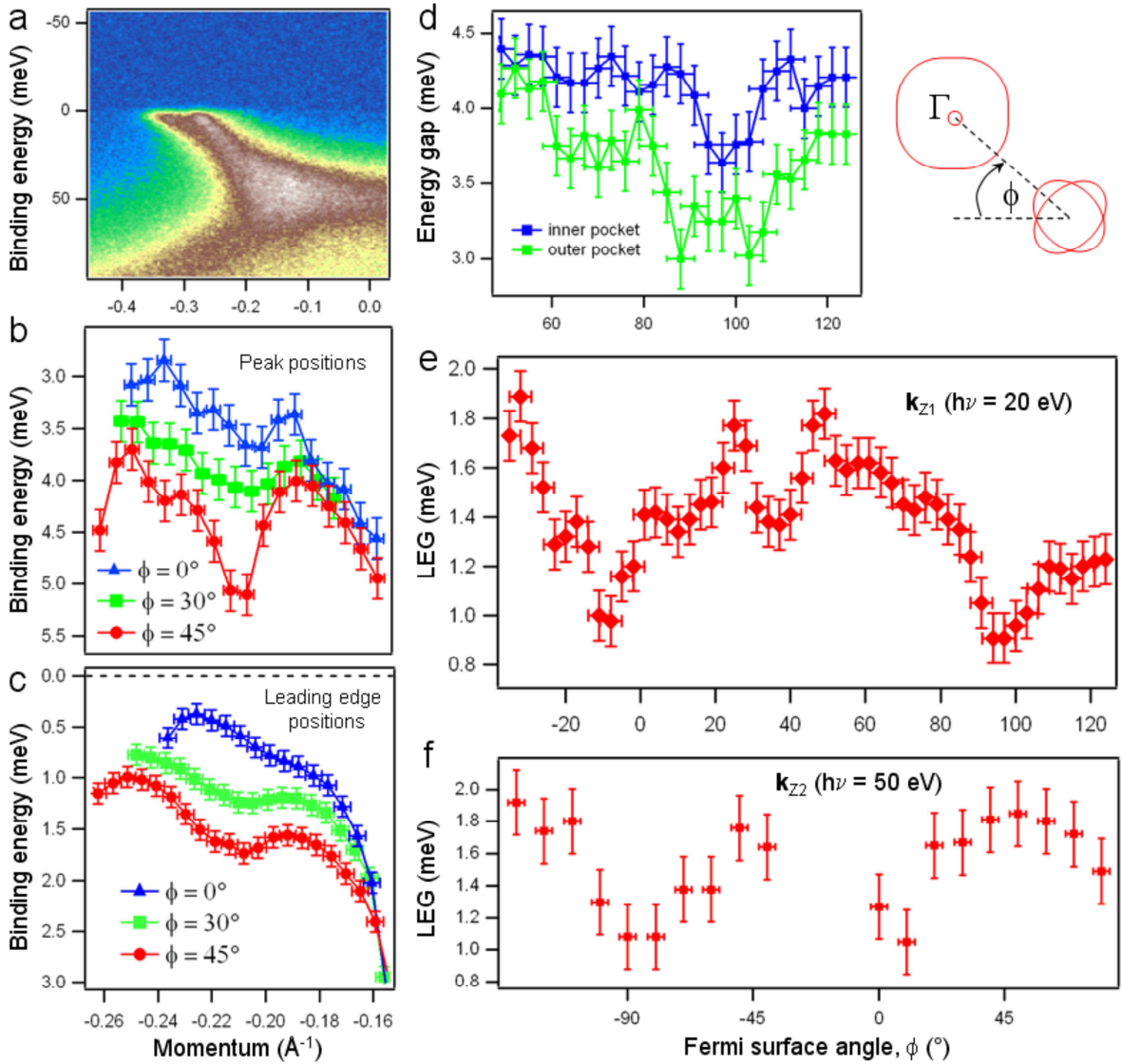

Figure 2. **Superconducting gap on the electron-like pockets. a)** Typical energy-momentum distribution along a radial momentum cut from the corner of the BZ showing two dispersing features which form electron-like FSs. Note, that two features appear to cross each other at ~ 30 meV leading to different orbital character of the small FS portion closest to $\Gamma$ than in the typical calculations: it is of $d_{xz,yz}$ character while the corresponding part of the inner pocket is $d_{xy}$. **b)** Peak positions for several radial cuts. Local maxima correspond to $k_F$-EDCs and their binding energies are the energy gaps. **c)** Maximal derivative or leading edge positions for the same cuts as in b. **d)** Gap functions in a limited angular interval showing in-phase variation of the gaps on both FSs. **e)** LEG from integrated EDCs obtained by summing the spectral weight along the radial cuts, as e.g. between -0.45 Å$^{-1}$ and 0.0 Å$^{-1}$ in panel a. **f)** Same as in e), but for different $k_z$, set by different photon energy. Angle $\phi$ is defined in the inset. Since the momentum separation between two electron-like FSs along the borders of the BZ is negligible one can consider them either as crossed ellipses or as inner and outer pockets of the same topology.

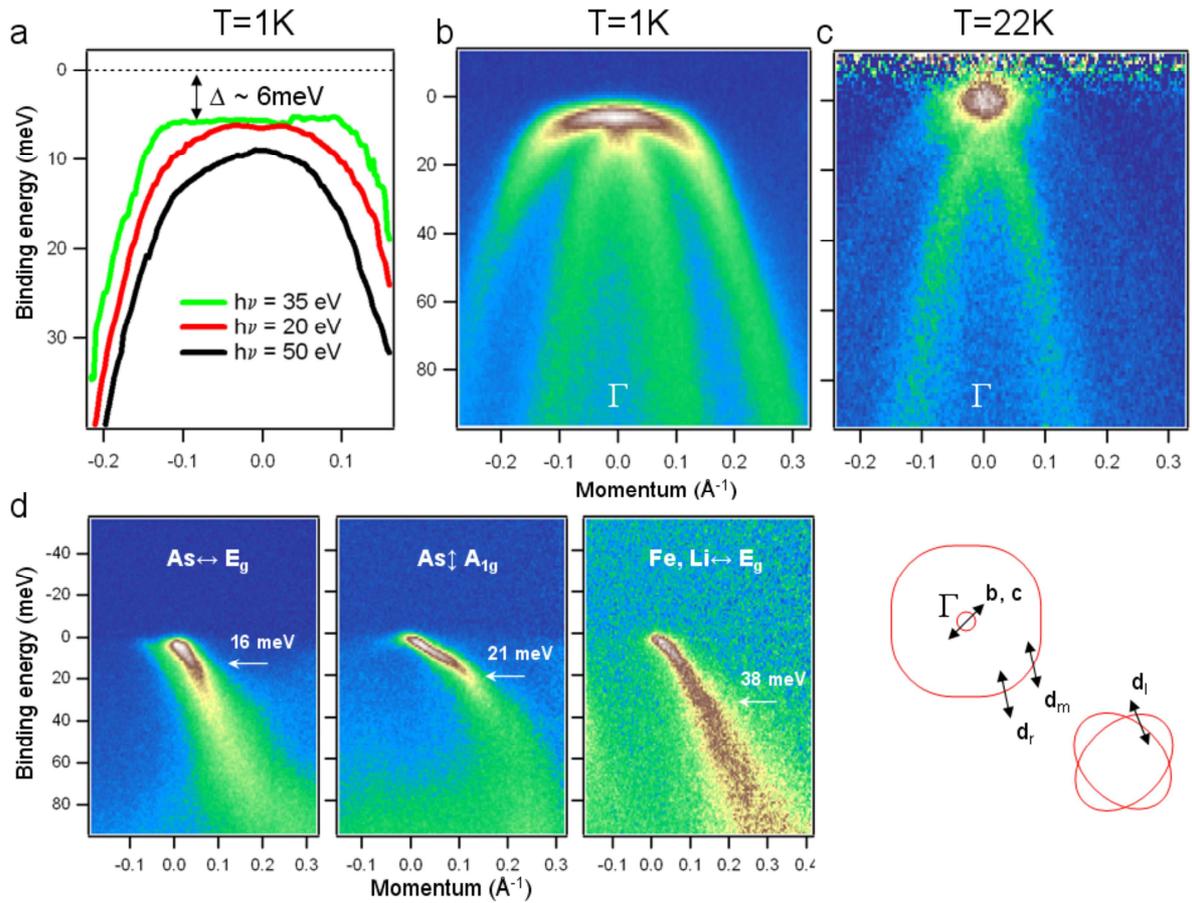

Figure 3. **Gap at Γ and kinks. a)** Dispersion of the middle hole-like feature near Γ for different $k_z$, set by various excitation energies. **b,c)** Energy-momentum distributions taken below and above $T_c$ for the case when the middle band crosses $E_F$. It is also seen that even in this case the inner hole-like feature does not reach the Fermi level. d) Kinks in the dispersion at different locations in the BZ seen as abrupt changes in the slope of the dispersing features (hν=20 eV). Matching phonon modes are indicated as well. Inset shows the momentum location of the cuts from b), c) and left/middle/right panels of d).

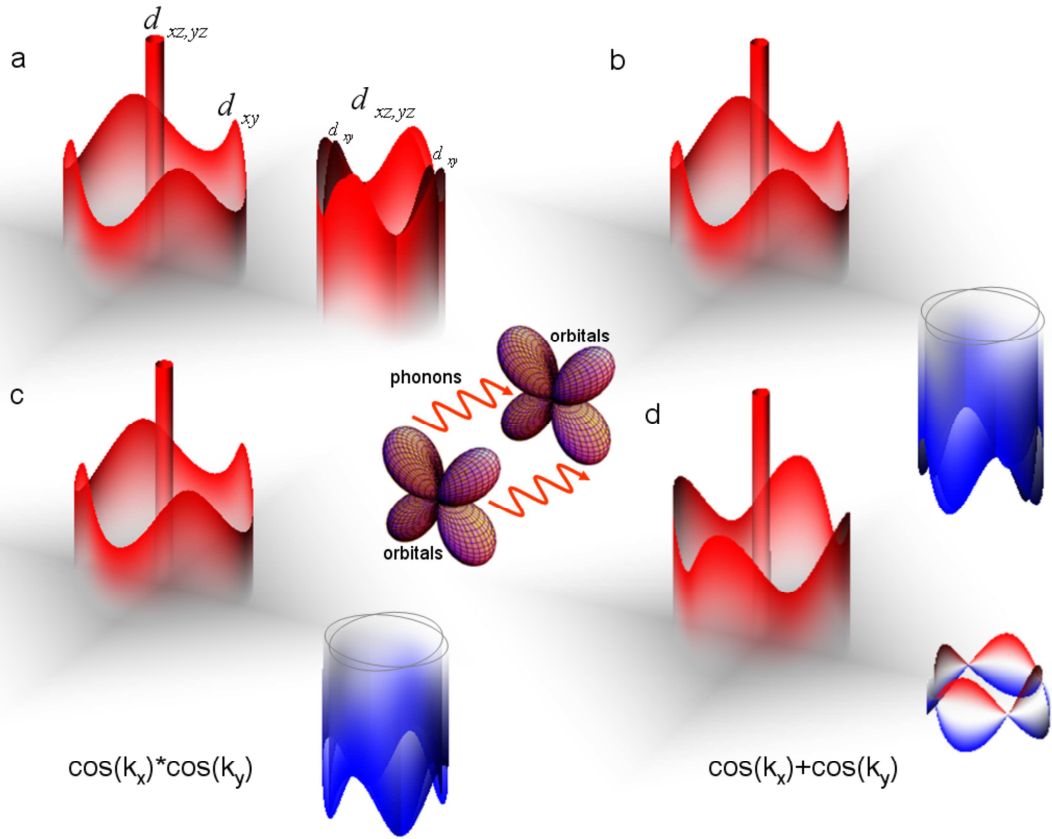

Figure 4. **Symmetry of the order parameter in LiFeAs. a)** 3D representation of the gap function as it follows from ARPES measurements. Only two sets of the Fermi surfaces are shown: those centered at Γ-point and at the corner of the BZ. Note the orientation of the extrema of all FSs and in-phase behavior of the electron-like sheets. Orbital composition of the FSs is indicated as well. Because of the crossing of the dispersions (Fig. 2a) the $d_{xy}$ character corresponds to the small portions of the *inner* electron pockets unlike in most calculations. **b)** Same as a), but taking into account the possibility of different sign between hole- and electron-like FSs. **c)** Gap function corresponding to the nodeless s± order parameter, usually discussed for hole-doped IBSC with nested FSs. Though similar to the case shown in b), the orientation of the extrema on the electron-like FSs is different. **d)** Nodal s± order parameter which should apply to LiFeAs according to the calculations within the spin-fluctuations scenario. Inset schematically represents the orbital-fluctuations model with electron-phonon interaction. Only the calculations based on this latter scenario qualitatively reproduce our experimental observations (see e.g. Ref. 14).

METHODS

**ARPES**

Photoemission experiments have been carried out using the synchrotron radiation from the BESSY (Helmholtz-Zentrum Berlin) storage ring. The end-station "1-cubed ARPES" is equipped with the He3 cryostat which allows collecting the angle-resolved spectra at temperatures below 1K. The overall energy resolution ranged from ~2.5 meV at hv = 15 eV to ~10 meV at hv =120 eV. All single crystals have been cleaved in UHV exposing the mirror-like surfaces.

**Single crystals**

Five large ( ~ 3x2x0.5mm) single crystals of LiFeAs ($T_c$=18 K) have been used in the current study, four Li11(SF) and one Li11(Sn). All operations on preparation of the large LiFeAs single crystals have been carried out in a dry box in Ar atmosphere. To grow the single crystal of LiFeAs from the melted tin (sample Li11(Sn)) the reagents in molar ratio Li:Fe:As:Sn=1.15:1:1:24 were mixed in alumina crucible which was inserted into Nb container sealed under 1.5 atm of argon gas. The Nb container was then sealed in an evacuated quartz ampoule and heated to 1163 K, and after that it was slowly cooled down to 853 K. At this temperature, the liquid tin was removed by decantation and then by centrifugation at high temperature. To grow the LiFeAs single crystals by self-flux method (Li11(SF)) the reagents in molar ratio Li:Fe:As = 3:2:3 were inserted to the same package heated up to 1363 K, kept at this temperature for 5 hours and cooled at a rate 4.5 K/h down to 873 K and then the furnace was switched on. The plate-like single crystals were separated from the flux mechanically. Phase identification was performed by means of X-Ray powder diffraction analysis of the polycrystalline samples, prepared from the single crystals by grinding them in a dry box. The results of tetragonal P4/nmm unit cell refinement (a=3.7701(15), c=6.3512(25) Å, V=90.27(8) Å$^3$ for Li11(SF) and a=3.7680(24), c=6.339(4)Å, V=90.00(13) Å$^3$ for Li11(Sn)) are in good agreement with the data available in the literature. The molar ratio of Fe:As close to 1:1, as well as the existence of about 0.5 mol. % of tin in the Li11(Sn) crystals have been identified from the EDX data.

**SUPPLEMENTARY INFORMATION**

Apart from the mentioned in the main text properties, LiFeAs possesses many other remarkable among IBSC characteristics: the residual resistivity ratio is one of the largest, the material crystallizes in a simple tetragonal lattice so that the unfolding of the BZ is not complicated, Fe-As-Fe angle is the sharpest and there is a glide plane between two Li layers resulting in a non-polar cleave. The absence of nesting and thus the replica because of magnetic folding, moderate three-dimensionality and simple orbital composition offer a unique opportunity to study the gap function in this material. The typical $\mathbf{k}_F$-EDC and its derivative are shown in Fig.S1 together with energy-momentum intensity plots. Well defined quasiparticles allowed us to determine the energy position of the peaks and leading edge midpoints with the precision of ~0.3 meV. The sensitivity of the latter to the superconductivity is demonstrated in Fig.S2.

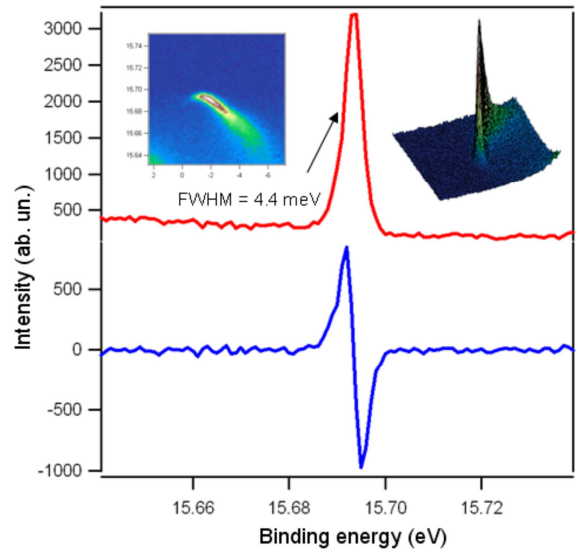

Fig.S1. Fermi momentum EDC (red) and its derivative (blue). Insets show corresponding energy-momentum distribution of the photoemission intensity.

In Fig. S3 we show the results of the fitting experimental data from Fig. 1e to the periodic function with higher harmonics with all three amplitudes left as free parameters. The obtained fitting function is $\Delta = 15.69257 - 0.00037*(\cos 4\phi + 0.12*\cos 8\phi - 0.2*\cos 12\phi)$ indicating the very significant contribution of the higher harmonics. Usually, the impurity scattering is responsible for a moderate contribution, but typical result would be the flattening of the maxima. In the present case the additional features are rather sharp, though not captured by the free parameters fit, and most likely are due to increased range of the pairing interaction. The observed deviations from $cos(4\phi)$ on the large $d_{xy}$ FS are indeed anomalous since reasonable gap functions would require that the expansion coefficients of the higher-harmonic basis functions decrease rapidly with increasing angular momentum.

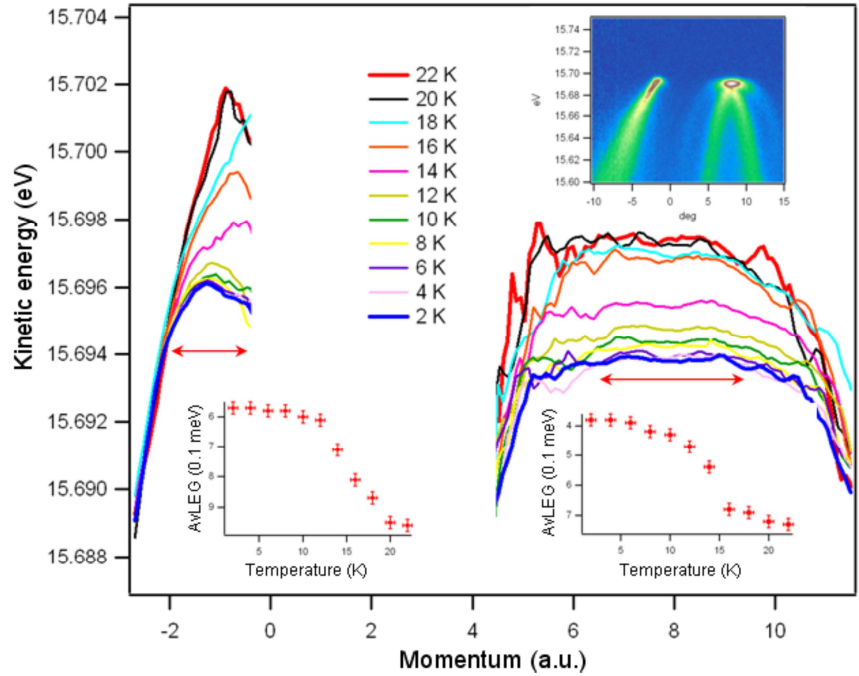

Fig.S2. Leading edge midpoint position as a function of temperature and momentum for the cut through all three hole-like features, as shown in inset. Temperature dependences of the average (over the momentum interval indicated by the double headed arrows) LE positions are shown as well.

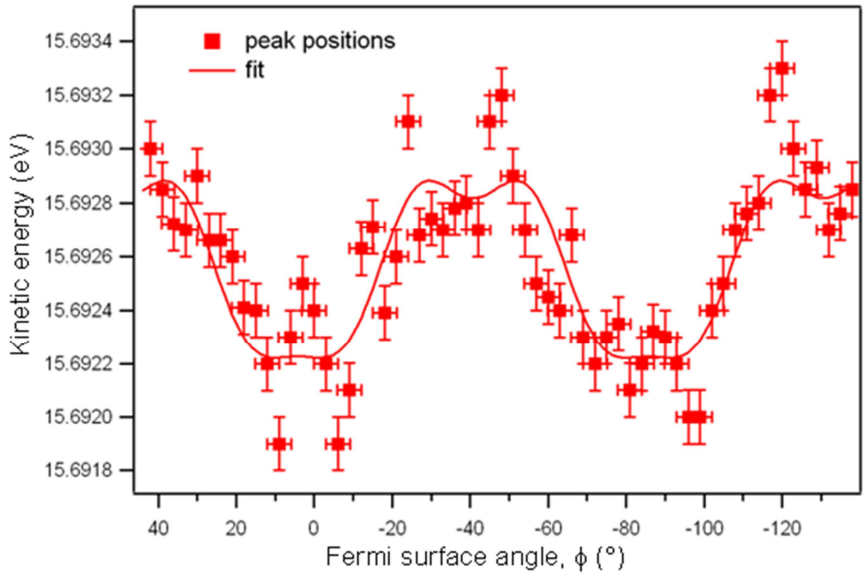

Fig.S3. Peak positions and fit.

One could, in principle, consider intraorbital $d_{xy}$ interaction as dominant since the maxima of the electron FS sheets' gaps are oriented towards center of the BZ, but then the orientation of the gap maxima on the large hole FS sheet would not match, instead pointing towards possible importance of the intraband scattering (with the coupling constant increased because of the nested character of the large hole pocket).

Two remarks should be made as regards the electron FS in the corner of the BZ. Because of the mutual crossing seen in Fig.2a to occur below the Fermi level, the orbital composition implied by our experiment differs from the one accepted in the literature. The small portions of the $d_{xy}$-character oriented towards the center correspond now to the inner electron FS, not to the outer. We note, however, that the separation of the electron-like FS to the inner and outer pockets itself is rather symbolic. The degeneracy along the sides of the BZ leading to the crossing of the ellipses from unfolded BZ is indeed lifted by the spin-orbit interaction, as well as the degeneracy of the middle and inner hole bands in the $\Gamma$-point, but the experimental splitting is at least an order of magnitude smaller than predicted in the calculations.

**Table S1** Phonon modes calculated (Ref. 24), measured at the center of BZ (Ref. 25) and ARPES kink energies from Ref. 10 and present study (in meV).

| meV | As ↔ $E_g$ | As ↕ $A_{1g}$ | Fe ↕ $B_{1g}$ | Fe ↔ $E_g$ | Li ↔ $E_g$ | Li ↕ $A_{1g}$ |
|---|---|---|---|---|---|---|
| **Theory (Ref.24)** | 15.1 | 23.5 | 28.1 | 30 | 36.8 | 44.5 |
| **Raman (Ref.25)** | - | 23 | 28-29 | 36 | 37-38 | 41-42 |
| **ARPES kinks (Ref.10,Fig.3d)** | 15, 16 | 21 | 30 | 38 | 38 | 44 |